\documentclass[a4paper]{jpconf}
\usepackage{graphicx}
\usepackage{citesort}
\usepackage{multirow}
\usepackage{capt-of}
\usepackage{amssymb}
\usepackage{lmodern}

\begin{document}
\title{Unbiased polarized PDFs\\ 
upgraded with new inclusive DIS data}

\author{Emanuele~R. Nocera}

\address{Rudolf Peierls Centre for Theoretical Physics, University of Oxford,
1 Keble Road, OX1 3NP Oxford, United Kingdom}

\ead{emanuele.nocera@physics.ox.ac.uk}

\begin{abstract}
I present a determination of longitudinally-polarized parton 
distribution functions of the proton from inclusive deep-inelastic
scattering data: {\tt NNPDFpol1.0+}. This determination, based on the
{\tt NNPDF} methodology, upgrades a previous analysis, {\tt NNPDFpol1.0},
in two respects: first, it includes all new data sets which have recently 
become available from the COMPASS experiment at CERN and from the E93-009, 
EG1-DVCS and E06-014 experiments at JLAB; second, it uses the state-of-the-art 
unpolarized parton set {\tt NNPDF3.0} as a baseline for the reconstruction
of fitted observables and for the determination of positivity constraints.
I discuss the impact of both these new inputs on the uncertainty of 
parton distribution functions. 
\end{abstract}

In the last years, the {\tt NNPDF} collaboration put a great deal of effort 
into the determination of minimally biased longitudinally-polarized
Parton Distribution Functions (PDFs) of the proton and their associated  
uncertainties. Two sets of longitudinally-polarized PDFs based on the 
{\tt NNPDF} methodology have been released so far: 
{\tt NNPDFpol1.0}~\cite{Ball:2013lla} and 
{\tt NNPDFpol1.1}~\cite{Nocera:2014gqa}. The former was obtained only from
inclusive Deep-Inelastic Scattering (DIS) data with longitudinally-polarized
beams and targets; the latter was obtained, in addition to the DIS data
included in {\tt NNPDFpol1.0}, also from open-charm production data by the 
COMPASS experiment at CERN and from $W^\pm$ and high-$p_T$ jet production data  
by the STAR and PHENIX experiments at RHIC. Respectively, RHIC data sets
have been proven to provide a first hint of flavor symmetry breaking for
polarized sea quarks, and a first evidence of a sizable, positive gluon
polarization in the proton~\cite{Nocera:2014gqa}.

Despite very promising, available RHIC data are not so many, and they cover a 
rather small kinematic range so far. Hence, the bulk of the experimental 
information on the longitudinally-polarized PDFs is still provided by 
inclusive DIS data: the leading observable, reconstructed from spin
asymmetries, {\it i.e.}, differences between cross sections 
with opposite target polarizations, is the polarized structure 
function of the nucleon $g_1(x,Q^2)$. Following factorization, this reads,
up to power-suppressed corrections, as
\begin{equation}
g_1(x,Q^2)
=
\frac{\sum_{q=1}^{n_f}e_q^2}{2n_f}
\left[
\mathcal{C}_{\rm NS}\otimes\Delta q_{\rm NS} + 
\mathcal{C}_{\rm S}\otimes\Delta\Sigma +
2n_f\mathcal{C}_g\otimes\Delta g
\right]
+
\frac{h^{\rm TMC}}{Q^2}
+
\frac{h}{Q^2}
+
\mathcal{O}\left(\frac{1}{Q^4}\right)
\,\mbox{,}
\label{eq:g1factorization}
\end{equation} 
where: $n_f$ is the number of active flavors; $e_q$ is the fractional electric 
charge for the $q^{\rm th}$ quark flavor; $\otimes$ denotes the usual convolution 
product; $\Delta q_{\rm NS}\equiv \sum_q^{n_f}\left(n_fe_q^2/\sum_{q=1}^{n_f}e_%q^2 
- 1 \right)\left(\Delta q + \Delta\bar{q}\right)$,
$\Delta\Sigma=\sum_{q=1}^{n_f}\left(\Delta q + \Delta\bar{q}\right)$ and 
$\Delta g$ are the nonsinglet, singlet and gluon longitudinally-polarized
PDFs; $\mathcal{C}_{\rm NS}$, 
$\mathcal{C}_{\rm S}$ and $\mathcal{C}_g$ are the corresponding leading-twist 
coefficient functions; and $h^{\rm TMC}$ and $h$ denote respectively (kinematic) 
target-mass corrections (TMCs) and (dynamic) higher-twist corrections. 
In Eq.~(\ref{eq:g1factorization}), the dependence of PDFs, coefficient 
functions, TMCs and higher-twist terms on both the scaling variable $x$ and 
the energy $Q^2$ has been omitted for brevity.

In this write-up, I present {\tt NNPDFpol1.0+}, an update of the analysis 
in Ref.~\cite{Ball:2013lla}, in which:
\begin{itemize}
 \item new data sets for the polarized structure function of the proton $g_1^p$ 
 from COMPASS (COMPASS-P15~\cite{Adolph:2015saz}), for the ratio of polarized 
 to unpolarized structure functions of the proton and deuteron 
 $g_1^{p,d}/F_1^{p,d}$ from CLAS 
 (JLAB-E93-009~\cite{Dharmawardane:2006zd,Guler:2015hsw} and
 JLAB-EG1-DVCS~\cite{Prok:2014ltt}), and for the virtual photoabsorption 
 asymmetry of the neutron $A_1^n$ from HALL-A 
 (JLAB-E06-014~\cite{Parno:2014xzb}) are fitted;
 \item the unpolarized PDF set used as a baseline both for the reconstruction 
 of the structure function $g_1(x,Q^2)$ from experimental asymmetries and 
 for the determination of positivity constraints (see respectively Secs.~2.1 
 and 4.4 in 
 Ref.~\cite{Ball:2013lla}) is updated from {\tt NNPDF2.1}~\cite{Ball:2011mu}
 to {\tt NNPDF3.0}~\cite{Ball:2014uwa}; the unpolarized structure
 functions, if needed, are obtained with {\tt APFEL}~\cite{Bertone:2013vaa}.
\end{itemize}
I will concentrate only on DIS data here; the {\tt NNPDFpol1.0+} 
parton set could then be reweighted with proton-proton collision data from 
RHIC along the lines of the analysis of Ref.~\cite{Nocera:2014gqa}. However,
this will require an additional study which is beyond the scope of this 
write-up.

\begin{figure}[t]
\centering
\begin{minipage}[t]{.45\textwidth}
\vspace{1.5pt}
\centering
\caption{\label{fig:kincov} Kinematic coverage of data.}
\includegraphics[scale=0.38, clip=true, trim=0cm 0cm 1.5cm 1.5cm]{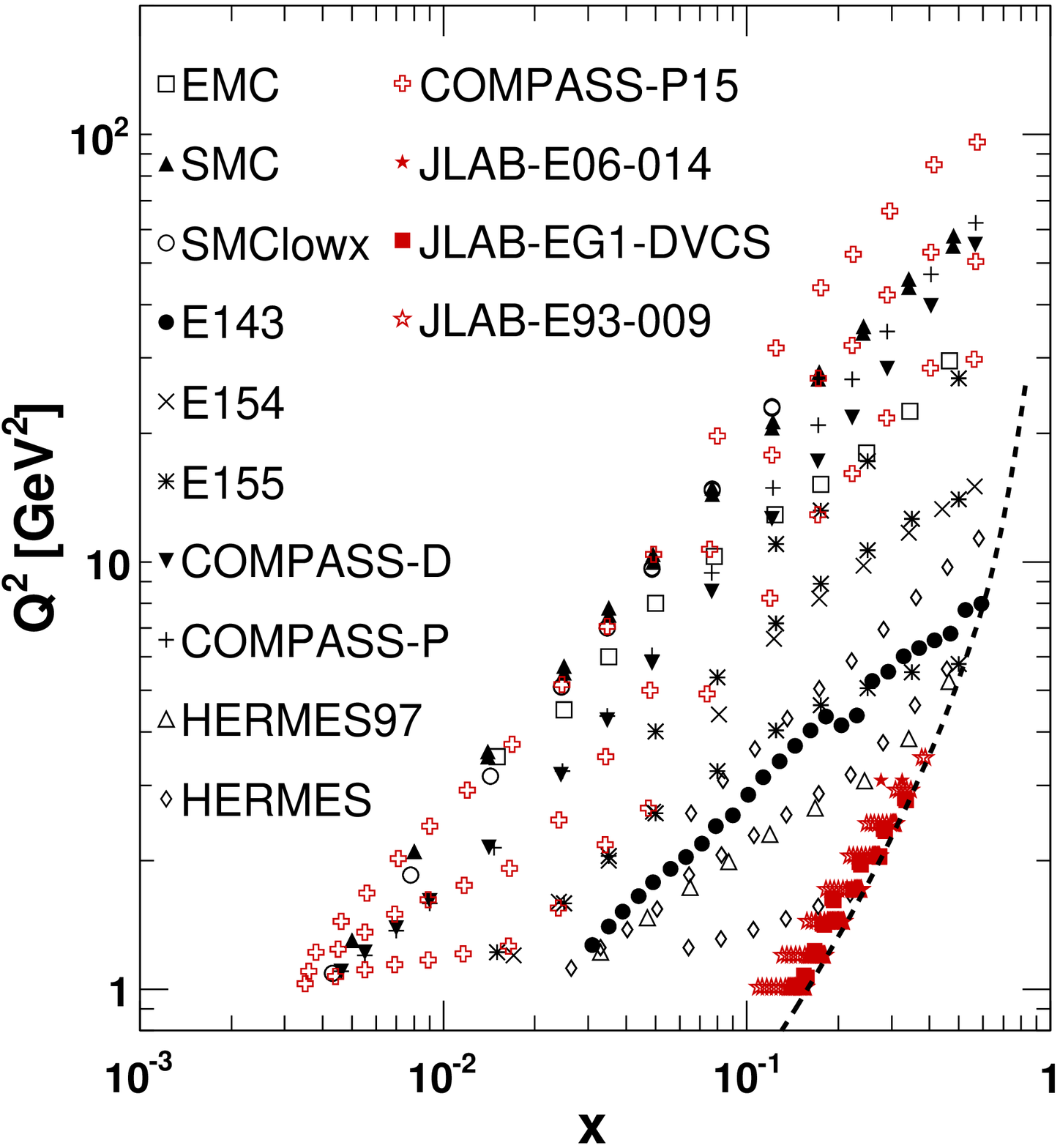}
\end{minipage}\hfill
\begin{minipage}[t]{.54\textwidth}
\centering
\captionof{table}{\label{tab:chi2val} 
Values of $\chi^2/N_{\rm dat}$ for each data set.}
 \footnotesize\rm
 \centering
 \begin{tabular}{lrlll}
 \br
 \multirow{2}*{Experiment} &
 \multirow{2}*{$N_{\rm dat}$} &
 \multicolumn{3}{c}{$\chi^2/N_{\rm dat}$}\\
 & & \tt{1.0} & \tt{1.1} & \tt{1.0+}\\
 \mr
 EMC            &  10  & 0.44  & 0.43  & 0.43\\
 SMC            &  24  & 0.93  & 0.90  & 0.92\\
 SMClowx        &  16  & 0.97  & 0.97  & 0.94\\
 E143           &  50  & 0.64  & 0.67  & 0.63\\
 E154           &  11  & 0.40  & 0.45  & 0.34\\
 E155           &  40  & 0.89  & 0.85  & 0.98\\
 COMPASS-D      &  15  & 0.65  & 0.70  & 0.57\\
 COMPASS-P      &  15  & 1.31  & 1.38  & 0.93\\
 HERMES97       &   8  & 0.34  & 0.34  & 0.23\\
 HERMES         &  56  & 0.79  & 0.82  & 0.69\\
 \mr
 COMPASS-P15    &  51  & 0.68* & 0.69* & 0.65\\
 JLAB-E93-009   & 148  & 1.26* & 1.23* & 0.94\\
 JLAB-EG1-DVCS  &  18  & 0.45* & 0.59* & 0.29\\
 JLAB-E06-014   &   2  & 2.81* & 3.20* & 1.33\\
 \mr
 {\scriptsize * Experiment not included}              
                &      & 0.77  & 0.78  & 0.74\\
 \br
 \end{tabular}
\end{minipage}
\end{figure}

Except for the improvements listed above, the analysis presented here proceeds 
exactly as in Ref.~\cite{Ball:2013lla}. The kinematic coverage of experimental 
data is shown in figure~\ref{fig:kincov} (new data sets are listed in the right 
column), together with the kinematic cut 
$W^2=m^2+Q^2(1-x)/x\geq 6.25$ GeV$^2$, with $m$ the nucleon mass. 
This was chosen so that the dynamic higher twist $h$ in 
Eq.~(\ref{eq:g1factorization}) become compatible with zero once
fitted to experimental data, and can then be neglected; 
TMCs instead are included exactly, see Sec.~3.2 in Ref.~\cite{Ball:2013lla}
for details. The structure function $g_1(x,Q^2)$ is 
reconstructed from the measured observables according to the available 
experimental information: referring to Sec.~2.1 of Ref.~\cite{Ball:2013lla},
the COMPASS-P15 data set is treated as the EMC data set, JLAB-E93-009 and 
JLAB-EG1-DVCS as E155, and JLAB-E06-014 as E143.
%------------------------------------------------------------------------------
%------------------------------------------------------------------------------

The quality of the {\tt NNPDFpol1.0+} analysis is assesed by the values
of the $\chi^2$ per data point, $\chi^2/N_{\rm dat}$, which are reported in 
table~\ref{tab:chi2val} for each data set, together with the number of 
included data points $N_{\rm dat}$. 
In table~\ref{tab:chi2val}, I also show the values of $\chi^2/N_{\rm dat}$ 
obtained from the {\tt NNPDFpol1.0}~\cite{Ball:2013lla}
and {\tt NNPDFpol1.1}~\cite{Nocera:2014gqa} parton sets. 
%note that the latter is based, on top of the same inclusive DIS data 
%included in {\tt NNPDFpol1.0}, also on open-charm production data from COMPASS 
%and high-$p_T$ jet and $W^\pm$ production data from STAR and PHENIX, see 
%Ref.~\cite{Nocera:2014gqa} for details. 
Inspection of table~\ref{tab:chi2val} allows for the following remarks.
\begin{itemize}
 \item The quality of the {\tt NNPDFpol1.0+} PDF determination, as
 measured by its total $\chi^2$ per data point 
 ($\chi^2_{\rm tot}/N_{\rm dat}=0.74$), is good, and comparable to that 
 achieved in previous determinations, both {\tt NNPDFpol1.0}
 ($\chi^2_{\rm tot}/N_{\rm dat}=0.77$)
 and {\tt NNPDFpol1.1} ($\chi^2_{\rm tot}/N_{\rm dat}=0.78$). 
 \item In comparison to {\tt NNPDFpol1.0} and {\tt NNPDFpol1.1}, the value of
 $\chi^2/N_{\rm dat}$ for new experiments included in 
 {\tt NNPDFpol1.0+} improves substantially for all JLAB data sets, while
 the difference is less significant for the COMPASS data set. 
 This suggests that JLAB data are likely to have a sizable impact on PDFs, 
 while the effects of COMPASS data are expected to be moderate (see also 
 the discussion in the sequel). 
 \item The value of $\chi^2/N_{\rm dat}$ for single experiments 
 are often well below the optimal value $\chi^2/N_{\rm dat}\sim 1$. This is a 
 consequence of the lack of experimental information on correlations among
 systematics, which cannot be accounted for properly, and ostensibly lead to
 an overestimation of experimental uncertainties. 
\end{itemize}

%-------------------------------------------------------------------------------
\begin{figure}[t]
\centering
\includegraphics[scale=0.215,clip=true,trim=1cm 1cm 1cm 0.5cm]{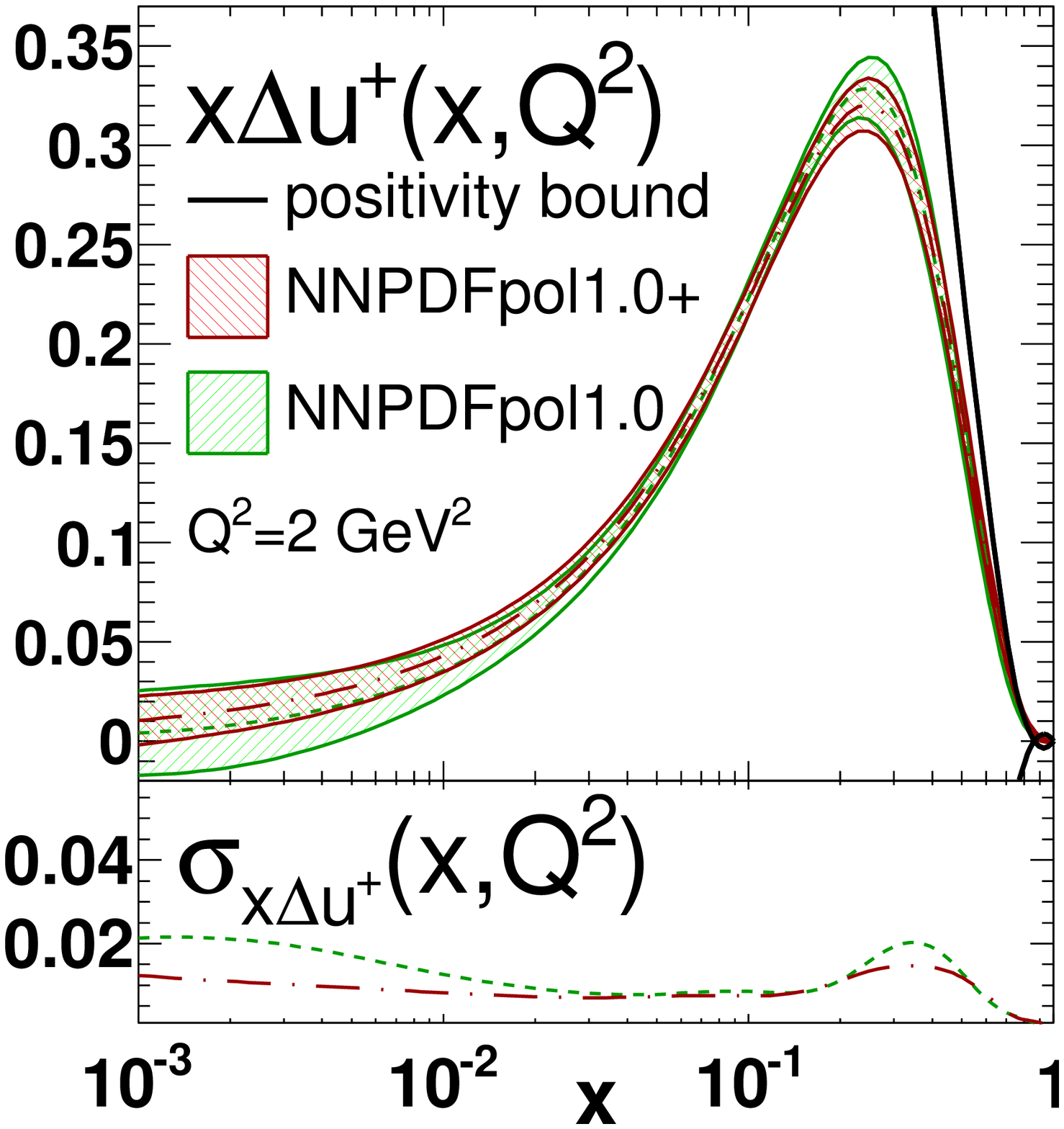}
\includegraphics[scale=0.215,clip=true,trim=1cm 1cm 1cm 0.5cm]{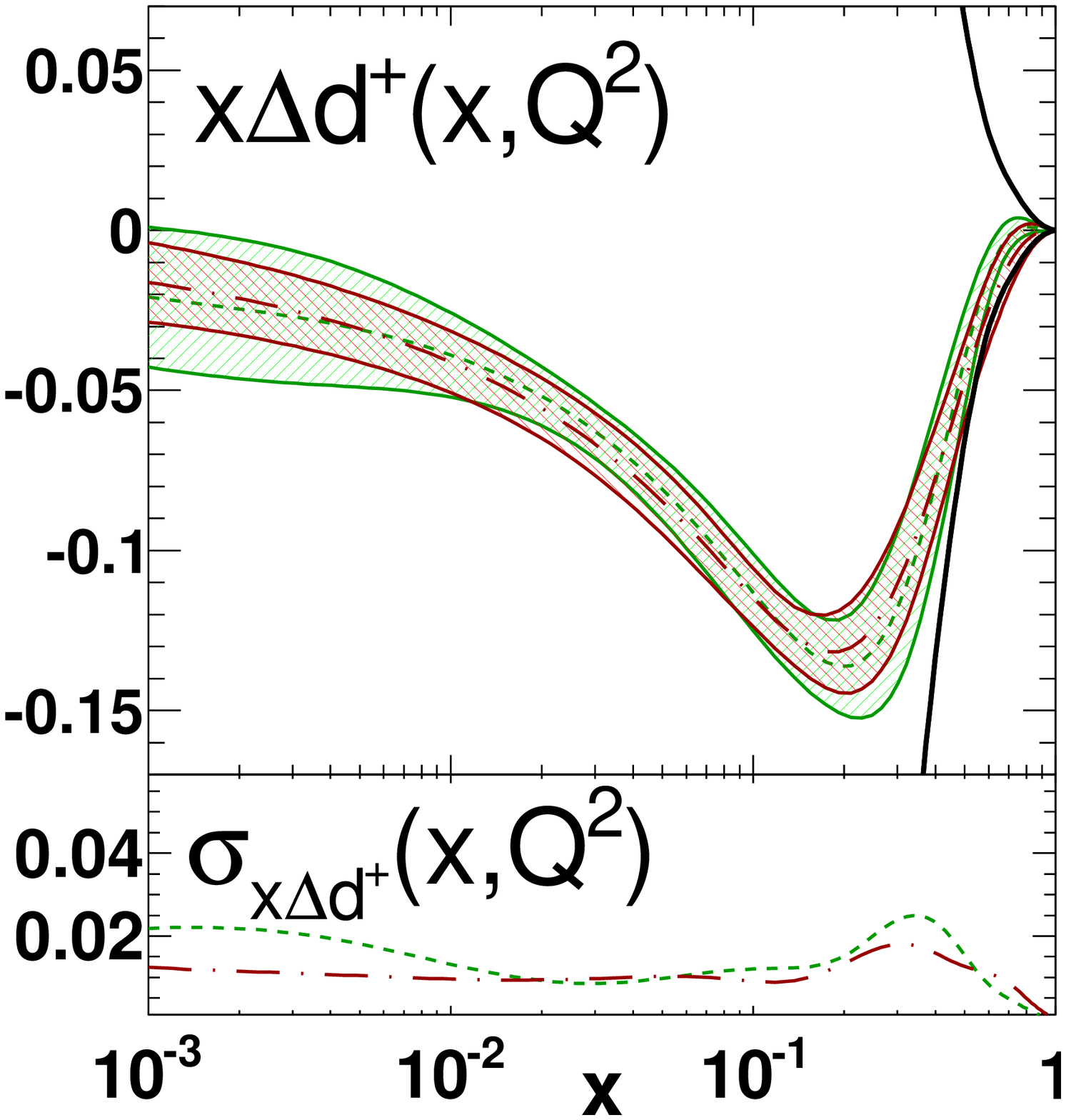}
\includegraphics[scale=0.215,clip=true,trim=1cm 1cm 1cm 0.5cm]{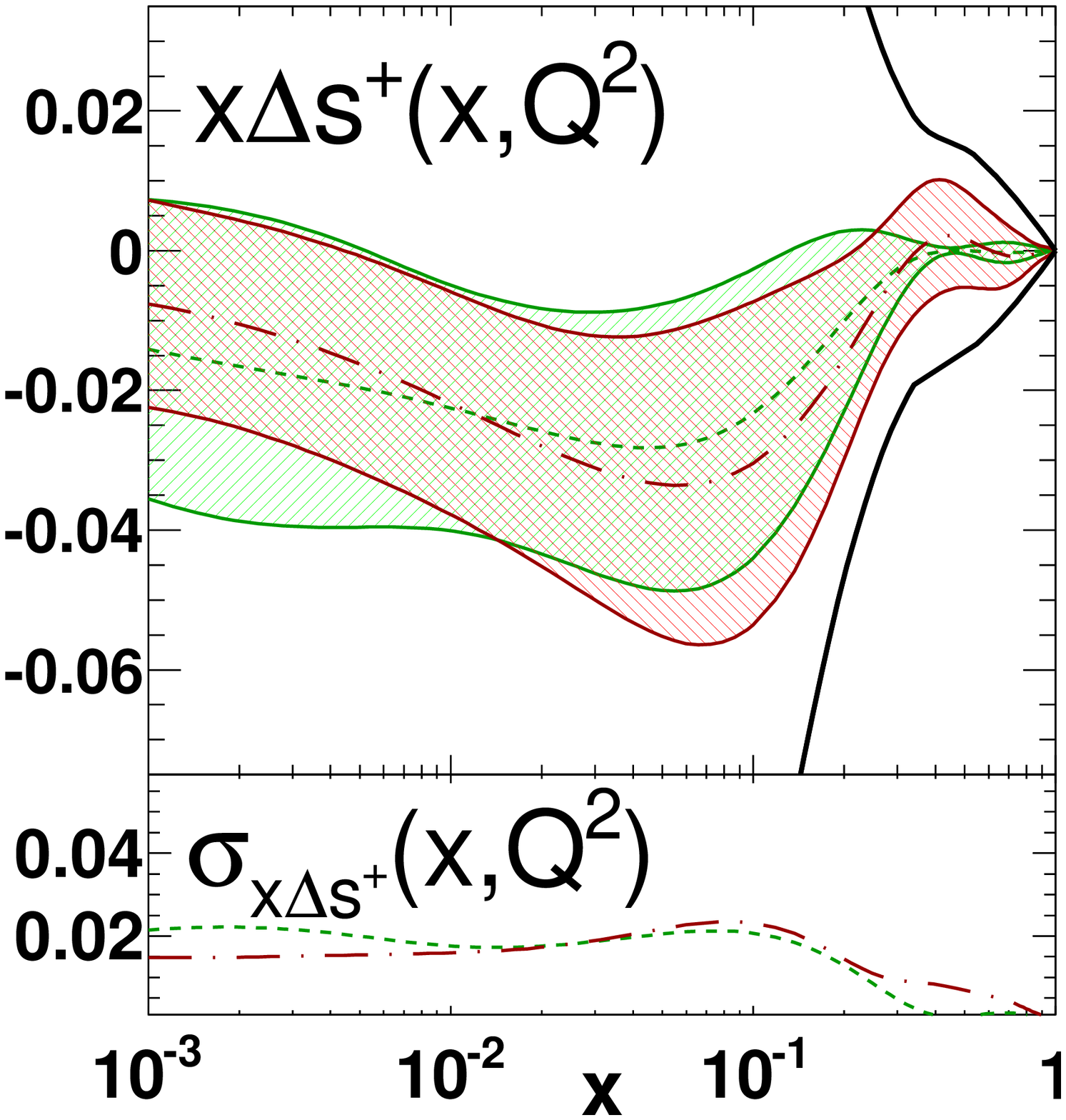}
\includegraphics[scale=0.215,clip=true,trim=1cm 1cm 1cm 0.5cm]{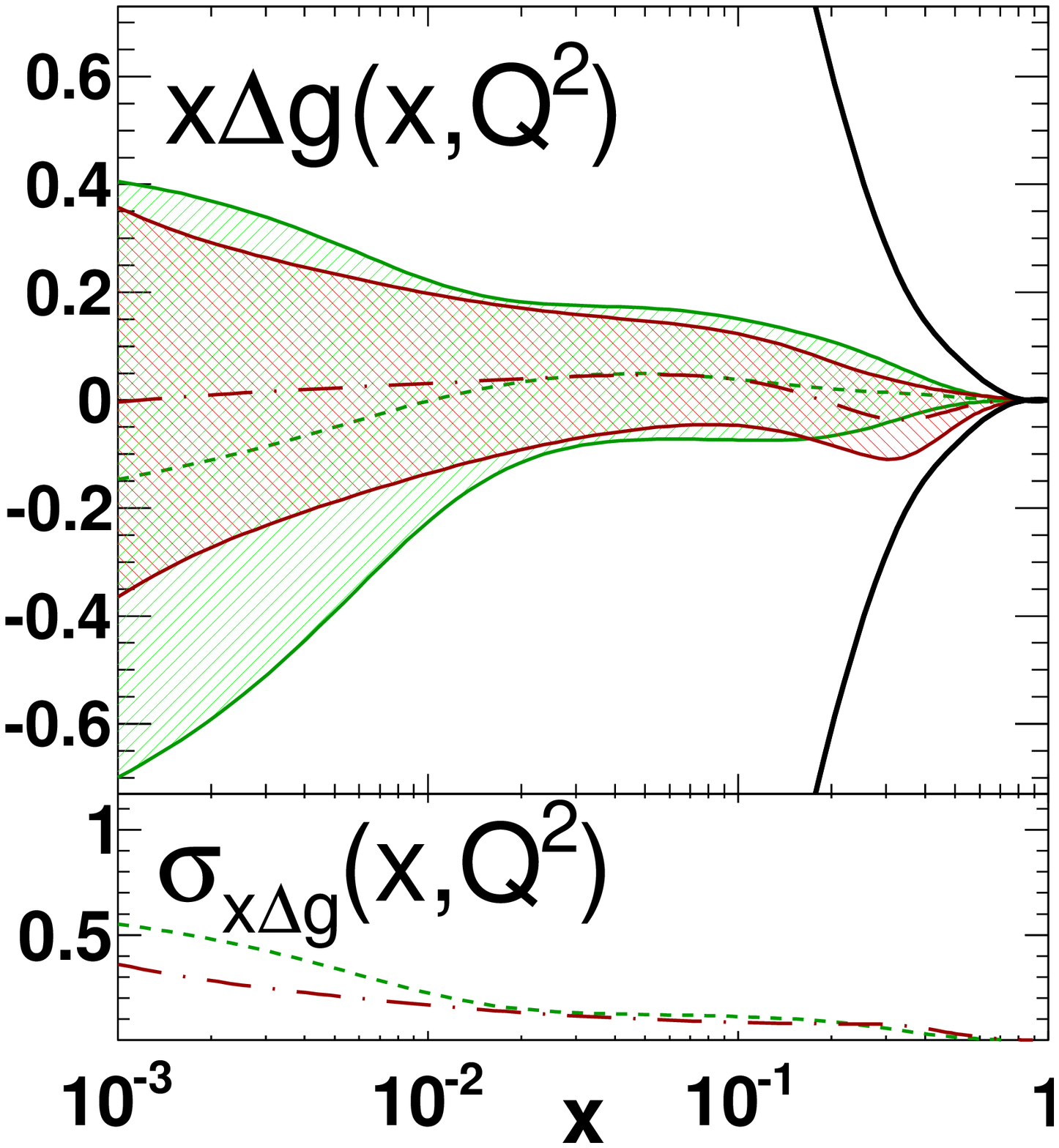}\\
\caption{Comparison of $x\Delta u^+$, $x\Delta d^+$, $x\Delta s^+$, $x\Delta g$,
and their uncertainties, from {\tt NNPDFpol1.0} and 
{\tt NNPDFpol1.0+} at $Q^2=2$ GeV$^2$. The positivity bound from 
{\tt NNDPF3.0} is also shown.}
\label{fig:PDFs}
\end{figure}
\begin{figure}[t]
\centering
\includegraphics[scale=0.215,clip=true,trim=1cm 1cm 1cm 0.5cm]{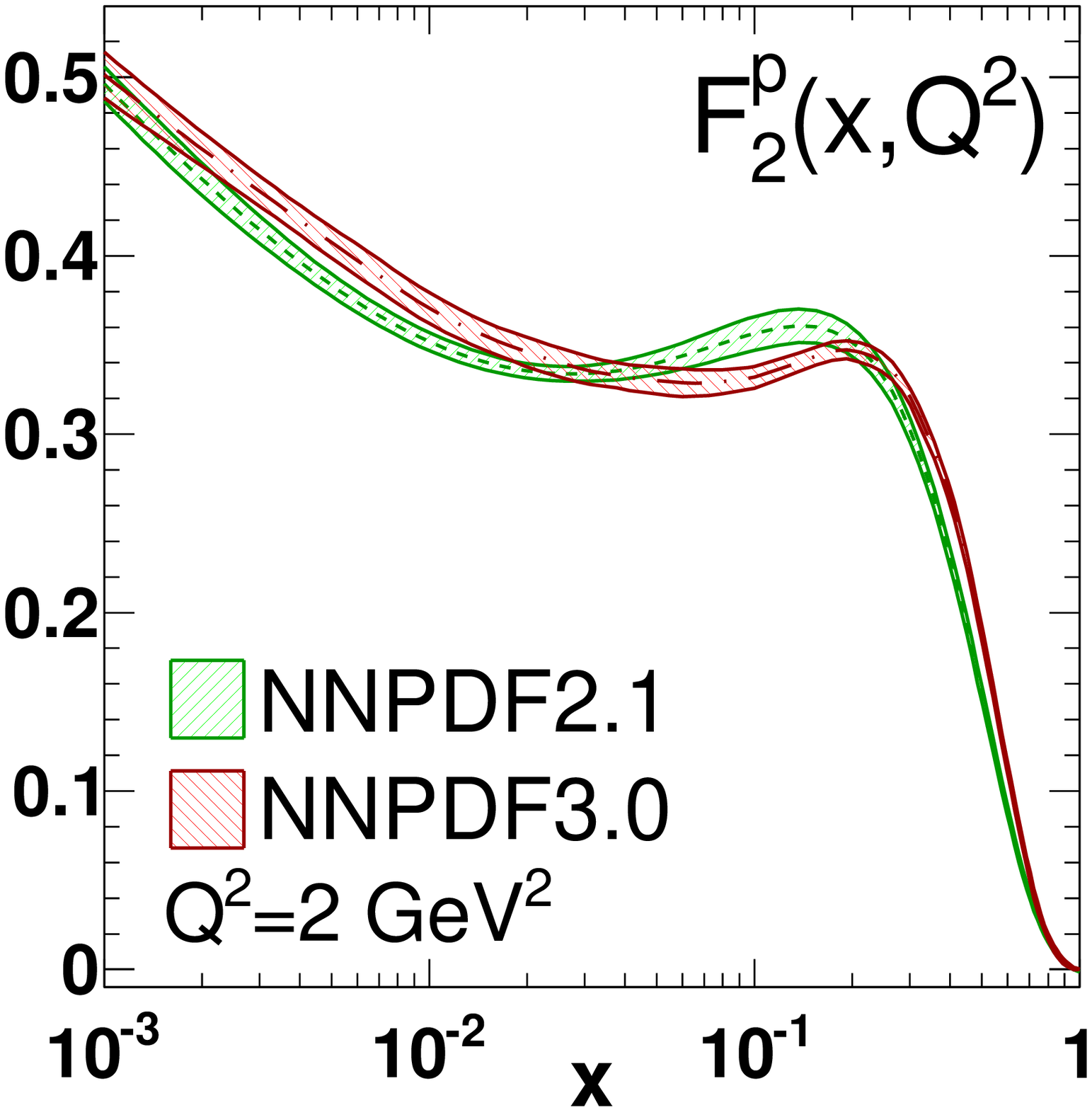}
\includegraphics[scale=0.215,clip=true,trim=1cm 1cm 1cm 0.5cm]{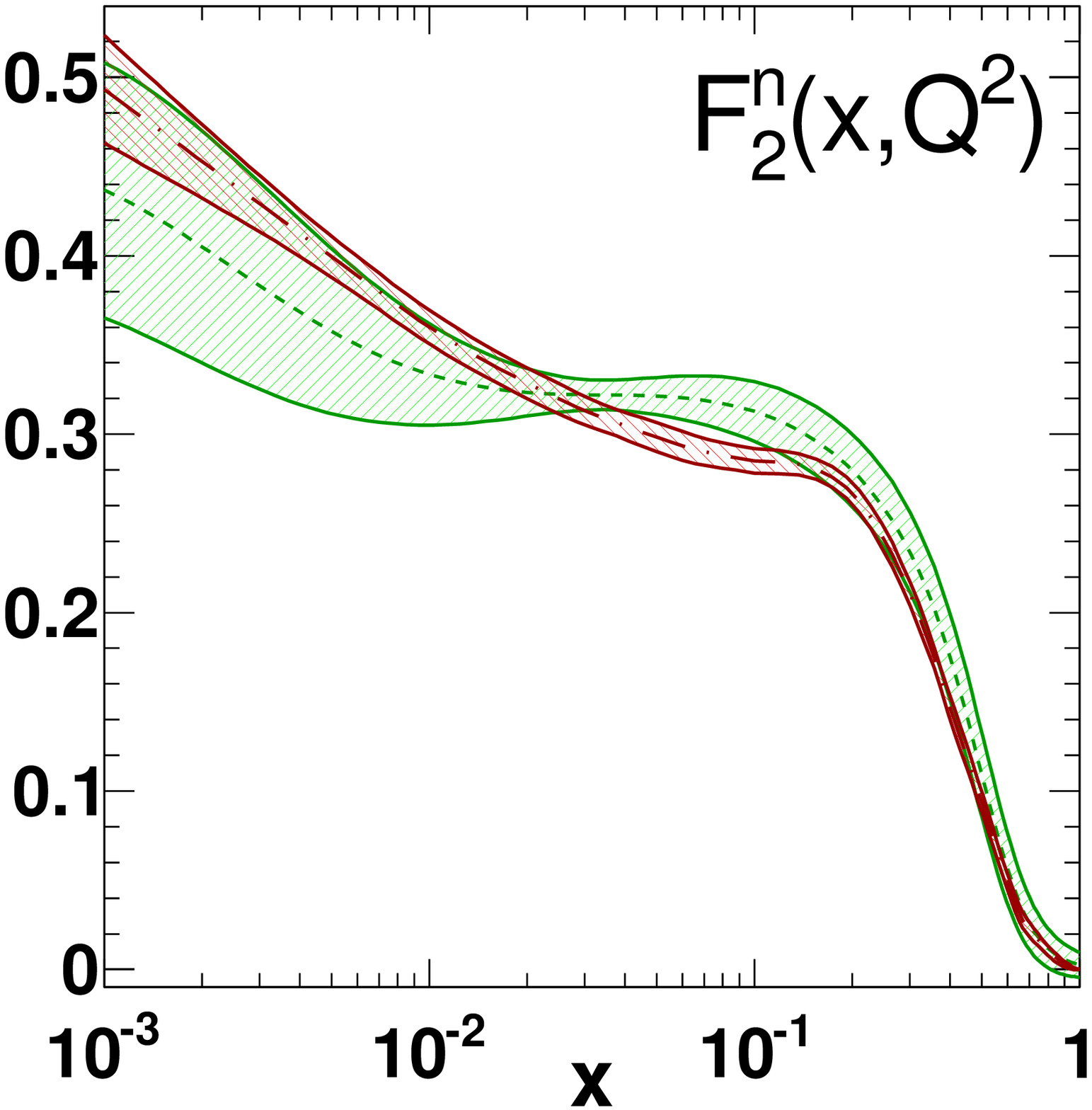}
\includegraphics[scale=0.215,clip=true,trim=1cm 1cm 1cm 0.5cm]{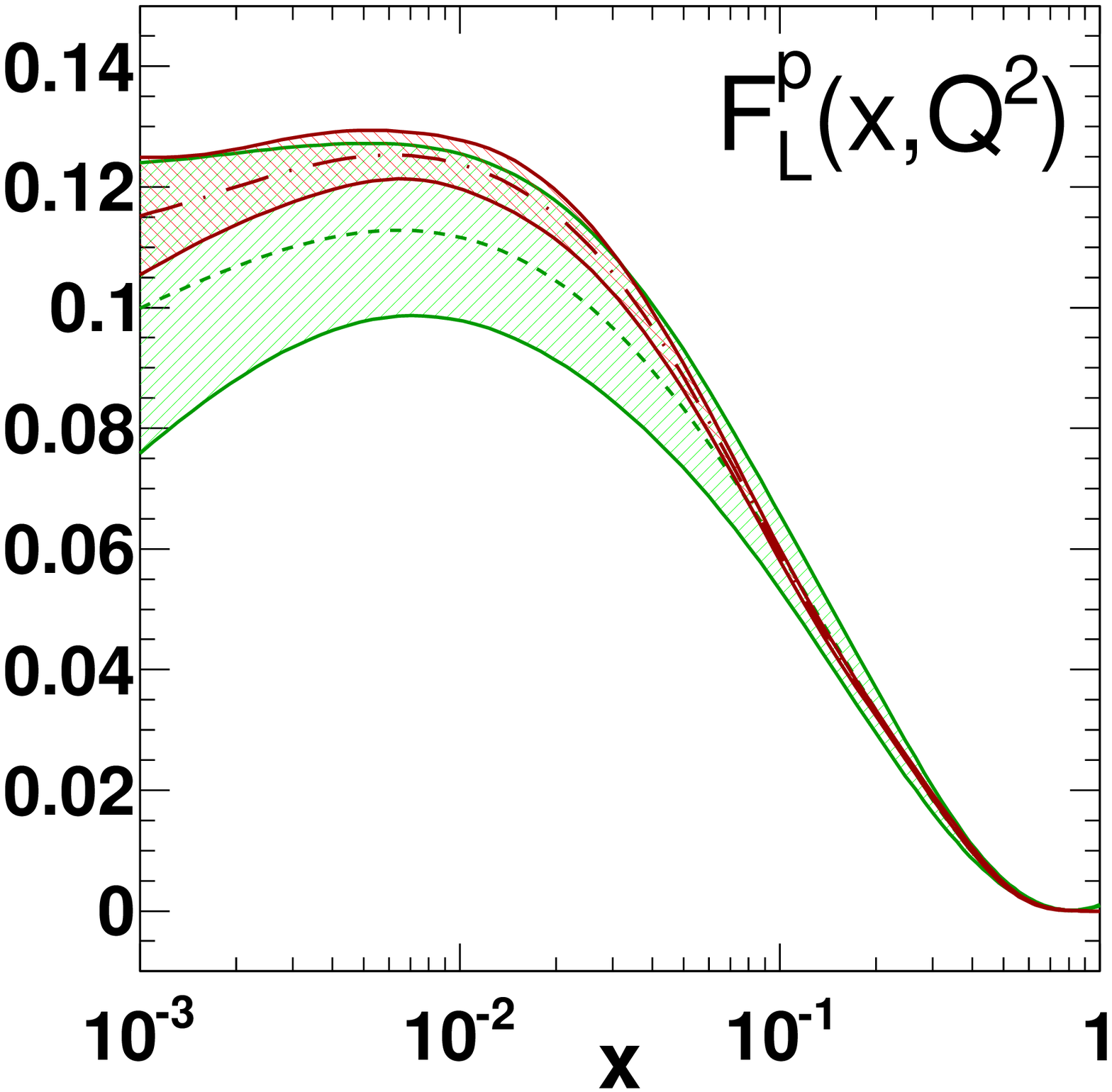}
\includegraphics[scale=0.215,clip=true,trim=1cm 1cm 1cm 0.5cm]{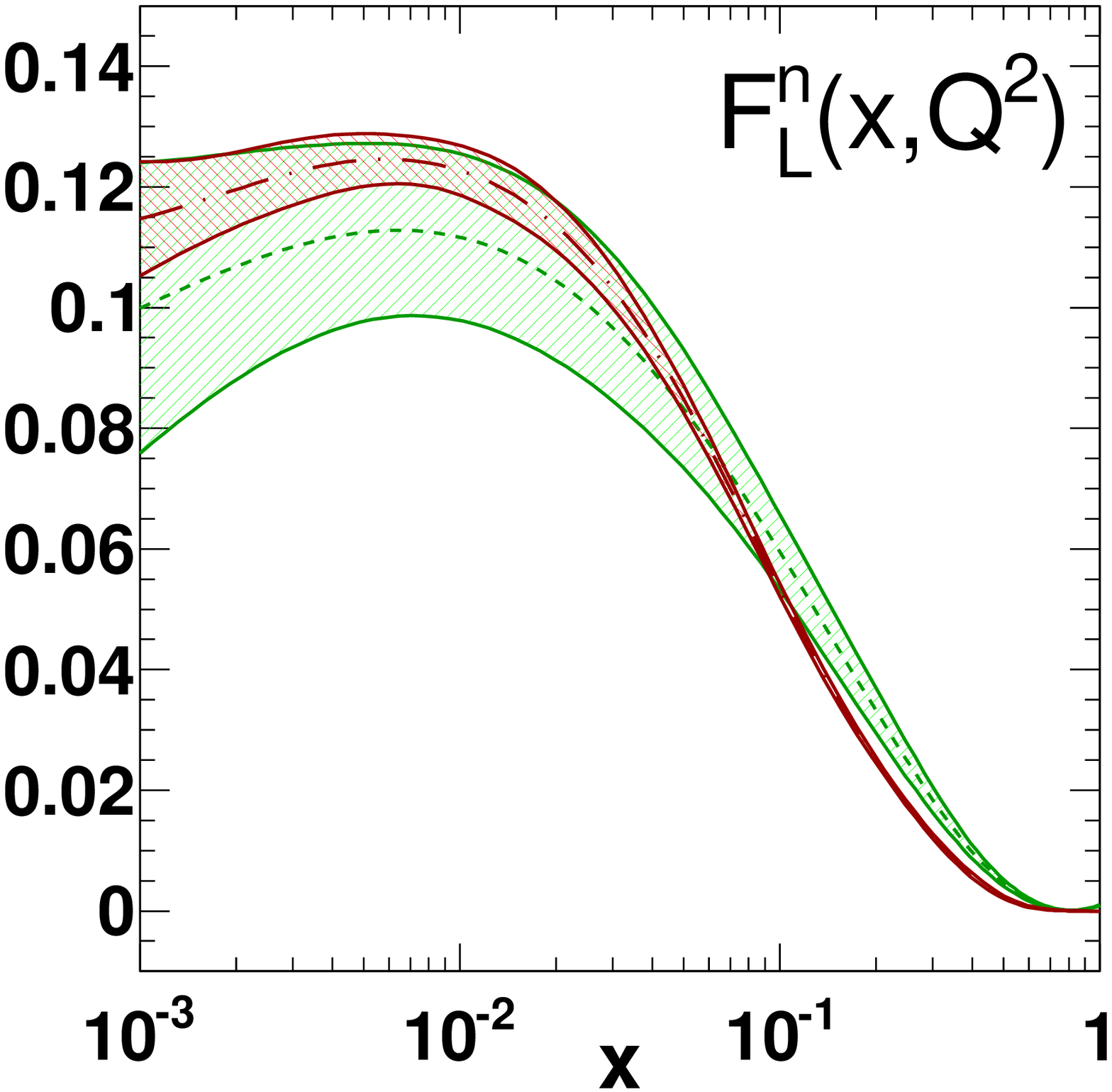}\\
\caption{The proton and neutron unpolarized structure functions
$F_2^{p,n}$ and $F_L^{p,n}$ from {\tt NNPDF2.1} 
and {\tt NNPDF3.0} at $Q^2=2$ GeV$^2$, as obtained with
{\tt APFEL}~\cite{Bertone:2013vaa}.}
\label{fig:unpSF}
\end{figure}

%------------------------------------------------------------------------------
In figure~\ref{fig:PDFs}, I show the total polarized quark distributions 
$\Delta q^+=\Delta q + \Delta\bar{q}$ ($q=u,d,s$) and the polarized gluon 
distribution $\Delta g$ at $Q^2=2$ GeV$^2$, obtained from both
{\tt NNPDFpol1.0+} and {\tt NNPDFpol1.0} PDF sets; their 
absolute uncertainties, $\sigma_{\Delta q}$, and the positivity bound
obtained from the {\tt NNPDF3.0} PDF set are also shown.
Inspection of figure~\ref{fig:PDFs} allows for the following remarks. 
%on the impact of both new data and the updated baseline unpolarized PDF set.
\begin{itemize}
 \item In the small-$x$ region ($x\lesssim 10^{-2}$), the uncertainty of all 
 quark and gluon distributions is reduced by about a factor two. 
 This is due to a better accuracy of the proton and neutron unpolarized 
 structure functions $F_2^{p,n}$ and $F_L^{p,n}$, whose uncertainty is propagated 
 to the polarized structure function $g_1^{p,n}$ when the latter is reconstructed
 from experimental asymmetries. The improvement between {\tt NNPDF2.1}
 (used in the original {\tt NNPDFpol1.0} analysis) and {\tt NNPDF3.0}
 (used in this analysis) is displayed in figure~\ref{fig:unpSF}, and is due to 
 the significant amount of new LHC data included in {\tt NNDPF3.0}. 
 The impact of COMPASS-P15, the only new data set 
 which covers the small-$x$ region, is instead rather limited: I performed
 a fit including COMPASS-P15 data, but using the same unpolarized structure 
 functions as in {\tt NNPDFpol1.0}, and I found that PDF uncertainties are 
 almost identical to those determined in {\tt NNPDFpol1.0}. 
 \item In the intermediate-to-large-$x$ region 
 ($10^{-2}\lesssim x \lesssim 0.6$), the uncertainty of $\Delta u^+$ and 
 $\Delta d^+$ is about two thirds of the uncertainty obtained
 from {\tt NNPDFpol1.0}. This is a genuine effect of new JLAB data sets, 
 which attain slightly larger $x$ values and are more accurate than those 
 included in {\tt NNPDFpol1.0}. The effects of updating the unpolarized 
 structure functions from {\tt NNPDF2.1} to {\tt NNPDF3.0}
 are less prominent than in the small-$x$ region: I checked explicitly
 that results are almost unchanged in this region if a fit including new data 
 but old unpolarized structure functions from {\tt NNPDF2.1} is performed.
 Note that the full potential of JLAB data, which includes the possibility to 
 discriminate among different non-perturbative models of nucleon 
 stucture~\cite{Nocera:2014uea}, could be exploited only by lowering the 
 kinematic cut on $W^2$. However, this will require a systematic inclusion of 
 dynamic higher-twist corrections: I  
 checked that, if $h$-terms in Eq.~(\ref{eq:g1factorization}) are neglected
 but $W^2\geq 4$ GeV$^2$ is taken, in principle the PDF uncertainty on 
 $\Delta u^+$ and $\Delta d^+$ can be reduced even more; however, the 
 value of the $\chi^2$ deteriorates significantly and PDFs become no longer
 reliable.
 \item In the large-$x$ region ($x\gtrsim 0.6$), no experimental data are 
 presently available to determine longitudinally-polarized PDFs. However, 
 their behavior is explicitly bounded by their unpolarized counterparts,
 as it follows from positivity constraints: at leading-order, 
 $|\Delta f(x,Q^2)|\leq f(x,Q^2)$, $f=u^+, d^+, s^+, g$. No significant 
 differences are found in the polarized PDFs by updating the baseline 
 unpolarized PDF set from {\tt NNPDF2.1} to {\tt NNPDF3.0}, except for 
 $\Delta s^+$. For this distribution, the uncertainty is about ten times 
 larger in {\tt NNPDFpol1.0+} than in {\tt NNPDFpol1.0}. 
 Note that $W+c$ data sets, sensitive to the 
 unpolarized strange distribution $s^+$, have been included in {\tt NNPDF3.0},
 which lead to a larger $s^+$ than that found in {\tt NNPDF2.1}. For
 this reason, the positivity bound on $\Delta s^+$ becomes less stringent, 
 and, in absence of any experimental information, this allows the uncertainty 
 on $\Delta s^+$ to grow significantly. A similar effect, though less prominent,
 can be noticed also for $\Delta g$.
\end{itemize} 
%------------------------------------------------------------------------------
%------------------------------------------------------------------------------

In conclusion, longitudinally-polarized PDFs obtained in this analysis
are well compatible with those obtained in the previous {\tt NNPDF} 
determination based on DIS data, {\tt NNPDFpol1.0}, but have slightly
smaller uncertainties. I have explicitly shown that there is a sizable 
interplay between unpolarized and polarized PDFs, and that JLAB data are 
effective in unveiling the large-$x$ behavior of PDFs; however, 
a determination of higher-twist corrections to $g_1$, 
Eq.~(\ref{eq:g1factorization}), will be mandatory to exploit their full 
potential. In the future, an extensive study will be dedicated to 
a global determination of longitudinally-polarized PDFs, into which the 
new DIS data discussed here, the proton-proton collision data 
included in {\tt NNPDFpol1.1} and possibly other new data  
will be incorporated together.

\section*{References}
\bibliography{Nocera-DSPIN15.bib}

\end{document}